\documentstyle[preprint,aps,epsf]{revtex}
\begin{document}
\draft
\begin{center}
{\Large Damped $\sin(\beta-\alpha)$ of Higgs couplings and the
lightest Higgs production at $\gamma\gamma$ colliders
in MSSM}\\
\end{center}
\vspace{1 cm}
\begin{center}
{\large Chao-Shang Huang\footnote{E-mail:csh@itp.ac.cn}
and Xiao-Hong Wu\footnote{E-mail:wuxh@itp.ac.cn}} \\
\vglue 1cm
Institute of Theoretical Physics, Chinese Academy of Sciences,\\
P. O. Box 2735, Beijing 100080, P. R. China \\
\end{center}
\vglue 0.5cm
\date{Nov. 22, 2001}
\vskip 2.5cm

\begin{center}
\begin{minipage}{14cm}
\begin{center} Abstract \\
\end{center}
\vspace{5mm}
In the decoupling limit, $M^2_{A^0} \gg M^2_Z$, the heavy CP-even, CP-odd
and charged Higgs boson masses are nearly degenerate, $\sin(\beta-\alpha)$
approaches $1$, and the lightest CP-even Higgs boson almost displays
the same properties as the Standard Model Higgs boson. But the stop and sbottom 
sector can change this pattern through radiative corrections.
We find that there are parameter regions at small, moderate and large $\tan\beta$ 
in MSSM under experimental constraints of $(g-2)_{\mu}$ and $b\rightarrow s\gamma$,
where $\sin^2(\beta-\alpha)$ is damped (say below 0.8),
which has a significant effect on Higgs couplings $g_{h^0VV} (V=W^\pm,Z^0)$
and $g_{h^0\gamma\gamma}$. We discuss its impact on the lightest CP-even Higgs
production at $\gamma\gamma$ colliders.
\end{minipage}
\end{center}

\newpage
\section{I\lowercase{ntroduction}}
The search for Higgs bosons and measurement of their properties are one of
the most important tasks at present and future colliders.
The Minimal Supersymmetry Standard Model(MSSM) has five physical
Higgs bosons, two CP-even Higgs $h^0$ and $H^0$, a CP-odd Higgs $A^0$ and
a charged Higgs pair $(H^+,H^-)$~\cite{mssm}. At tree level, the entire Higgs
mass spectra and mixing angle $\alpha$ of CP-even Higgs are determined
by only two independent parameters, conveniently chosen
CP-odd Higgs mass $M_{A^0}$ and the ratio
of two vacuum expectation values $\tan\beta$. And there
is an upper bound on the mass of lightest CP-even Higgs $h^0$,
$M_{h^0} < M_Z$. However, radiative corrections can change this bound.
At one loop level, $M^2_{h^0}$ receives the $G_FM^4_t$ term enhancement, which
shifts $M^{max}_{h^0}$ to $130$GeV~\cite{hmax}.

In the limit $M^2_{A^0} \gg M^2_Z$ (for $M_{A^0} \geq 300$GeV), the
charged, heavy CP-even and CP-odd Higgs bosons are nearly mass
degenerate, $M_{H^\pm} \simeq M_{H^0} \simeq M_{A^0}$, $\sin(\beta-\alpha)$
approaches $1$, and the properties of the lightest CP-even Higgs boson $h^0$
are almost identical to those of the Standard Model (SM) Higgs boson $h^{sm}$.
This is known as the decoupling limit~\cite{decouple}. When $M_{A^0}$ is not
too large, i.e., far from the decoupling limit,
$h^0$ and $H^0$ mix severely and
$\sin(\beta-\alpha)$ is lifted from $1$~\cite{higgsreview}.

Recently, the study of Higgs boson productions at photon colliders
has been extensively carried out~\cite{photoncollider}.
Photon colliders have distinct advantages in searches for and measurements 
of neutral Higgs bosons. At $\gamma\gamma$ colliders, the Higgs boson can be produced in $s$-channel
resonance, via triangle loop with all the charged particles
which gives the unique opportunity to precisely measure
the properties of Higgs boson, mass, production and decay channel and
determine CP property and spin, parity~\cite{gunion93}.
And $\gamma\gamma$ colliders
give a chance to produce single heavy Higgs $H^0$, $A^0$, which
extends the mass reach. Compared with
$e^+e^-$ colliders, the heavy Higgs bosons are produced in pair, because the
$ZH^0$, $ZA^0$ channels are suppressed due to nearly zero $\cos(\beta-\alpha)$.
A simulation study of the production of the lightest Higgs boson and discovery
potential at $\gamma\gamma$ colliders has been given~\cite{simulation}.
 
To study the Higgs boson productions at photon colliders is essentially to examine the loop
induced coupling of Higgs bosons to photons  $g_{h^0\gamma\gamma}$.  The coupling $g_{h^0\gamma\gamma}$ 
had been discussed long ago\cite{h2gammaold},
recent revisit in the decoupling limit in MSSM
as well as the tow-photon decay of the SM-like Higgs boson
at photon colliders has been worked out~\cite{h2gamma}. Related with the decoupling limit, a question arises naturally.
That is, is there any region with $M^2_{A^0} \gg M^2_Z$ of the parameter space in MSSM where 
the decoupling limit can be relaxed ( i.e., there are significant mass differences among
 $M_{H^\pm}, M_{H^0}$ and $M_{A^0}$ and $\sin(\beta-\alpha)$ deviates from $1$ 
although $M^2_{A^0} \gg M^2_Z$ )? 
In this paper we would like to answer the question and discuss its phenomenological
implication on the light neutral Higgs boson production at photon colliders. It is shown 
 that the decoupling limit can be relaxed in some
 regions with $M^2_{A^0} \gg M^2_Z$ of the parameter space which are allowed by experiments
of $(g-2)_{\mu}$, $b\rightarrow s \gamma$ and lower bounds of sparticle masses,
due to the large off-diagonal scalar top and scalar
bottom mass matrix elements contributing to the Higgs sector by radiative corrections. In the regions,
the charged, heavy CP-even and CP-odd Higgs bosons are not mass degenerate, and
$\sin^2(\beta-\alpha)$ can be damped (say, below $0.8$) from $1$, the value in the 
decoupling limit, which has significant effect
to the couplings proportional to $\sin(\beta-\alpha)$,
most importantly $g_{h^0VV},(V=W^\pm,Z)$ and consequently $g_{h^0\gamma\gamma}$
(note that the dominant contribution to $g_{h^0\gamma\gamma}$ comes from $W^+W^-$ loop, 
which is proportional to $g_{h^0W^+W^-}$). The discovery potential of Higgs bosons in experiments 
relies heavily on these couplings.

The paper is organized as follows. In section II we analyze how the decoupling limit
can be relaxed when including the radiative corrections to the neutral Higgs boson
mass matrix. The experimental constraints on the parameter space in MSSM are discussed
in section III and the formula concerned with the light neutral Higgs boson production
at photon coliders are listed in section IV. We present our numerical results in section
V. Finally, in section VI we end up with concluding remarks.

\section{CP-\lowercase{even} H\lowercase{iggs mass matrix 
and $\sin^2(\beta-\alpha)$}}
The mass-squared matrix of two CP-even Higgs bosons $h^0$ and $H^0$ is
given by
\begin{equation}
\label{higgsmatrix}
M^2_{\cal H} =
\left(
\begin{array}{cc}
M^2_{11}&M^2_{12}\\
M^2_{12}&M^2_{22}
\end{array}
\right)
=
\left(
\begin{array}{cc}
M^2_{A^0} s^2_\beta + M^2_Z c^2_\beta + \delta M^2_{11}&
-(M^2_{A^0} + M^2_Z) s_\beta c_\beta + \delta M^2_{12}\\
-(M^2_{A^0} + M^2_Z) s_\beta c_\beta + \delta M^2_{12}&
M^2_{A^0} c^2_\beta + M^2_Z s^2_\beta + \delta M^2_{22}
\end{array}
\right)
\end{equation}
$c_\beta$,$s_\beta$ denote $\cos\beta$ and $\sin\beta$.
$\delta M^2_{ij} (i,j=1,2)$ are generated through radiative corrections from
loops of fermions and sfermions and proportional to the fermion Yukawa couplings
squared. The dominant contribution comes from the the third generation, which
has large fermion masses. The corresponding
top and bottom squark mass-squared matrix in
$(\tilde{q}_L,\tilde{q}_R)$ basis are expressed in MSSM as
\begin{equation}
M^2_{\tilde{t}} =
\left(
\begin{array}{cc}
M^2_{\tilde{q}} + M^2_t + (\frac{1}{2} - \frac{2}{3}\sin^2\theta_W)
M^2_Z\cos2\beta&
M_t(A_t - \mu \cot\beta)\\
M_t(A_t - \mu \cot\beta)&
M^2_{\tilde{t}_R} + M^2_t + \frac{2}{3}\sin^2\theta_WM^2_Z\cos2\beta\\
\end{array}
\right)
\end{equation}
\begin{equation}
M^2_{\tilde{b}} =
\left(
\begin{array}{cc}
M^2_{\tilde{q}} + M^2_b + (-\frac{1}{2} + \frac{1}{3}\sin^2\theta_W)
M^2_Z\cos2\beta&
M_b(A_b - \mu \tan\beta)\\
M_b(A_b - \mu \tan\beta)&
M^2_{\tilde{b}_R} + M^2_b - \frac{1}{3}\sin^2\theta_WM^2_Z\cos2\beta
\end{array}
\right)
\end{equation}
The radiative corrections to $\delta M^2_{ij} (i,j=1,2)$,
including dominant one-loop
top, bottom quark and top, bottom squark corrections and
two-loop leading logarithmic contributions are given in~\cite{higgsmass}.
In our numerical calculation we used the results
of radiative corrections to the CP-even Higgs mass matrix based on
ref.~\cite{higgsfull} that incorporates the one-loop effective
potential and two-loop leading-log contribution from
arbitrary off-diagonal stop and sbottom matrices.
In the limit of vanishing off-diagonal parameters $\mu$, $A_t$ and $A_b$,
$\delta M^2_{11}$ and $\delta M^2_{12}$ is $0$, and
$\delta M^2_{22} = \frac{3g^2}{8\pi^2 M_W^2}
\frac{M_t^4}{\sin^2\beta} \ln(1+\frac{M^2_{\tilde{q}}}{M_t^2})$,
which lifts the upper bound of lightest Higgs mass $M_{h^0}$
above $M_Z$~\cite{hmax}.

The neutral CP-even Higgs mass eigenvalues can be derived from
Eq. (\ref{higgsmatrix}) as
\begin{equation}
M_{h^0,H^0} = \frac{Tr M^2_{\cal H} \mp \sqrt{(TrM^2_{\cal H})^2 - 
4 \det M^2_{\cal H}}}{2}\nonumber\\
\end{equation}
where $Tr M^2_{\cal H} = M^2_{11} + M^2_{22}$,
$\det M^2_{\cal H} = M^2_{11}M^2_{22} - (M^2_{12})^2$.

The mixing angle $\alpha$ of CP-even Higgs bosons can be defined as
\begin{equation}
\tan 2\alpha = \frac{\sin 2\beta (M^2_{A^0} + M^2_Z) - 2\delta M^2_{12}}
{\cos 2\beta (M^2_{A^0} + M^2_Z) + \delta M^2_{22} - \delta M^2_{11}}
\hspace{15mm}(-\frac{\pi}{2}<\alpha<\frac{\pi}{2})
\label{tan2alpha}
\end{equation}
In the decoupling limit, all $\delta M^2_{ij} (i,j=1,2)$ are order of $M^2_Z$ or
smaller than that, then $\alpha \sim \beta - \frac{\pi}{2}$, and
$\sin(\beta-\alpha) \sim 1$. However, $\sin(\beta-\alpha)$ can deviate from $1$ in some cases.
In order to see this explicitly, let us look at the cases of  moderate and large $\tan\beta$.
For moderate and large $\tan\beta$ (say, $\tan\beta=10$),
$\cos 2\beta \simeq -1$ and $\sin 2\beta \simeq 2\cot\beta$.
Then, when the numerator is not too small (of order $M^2_Z$), and the
large $-M^2_{A^0}$ term in the denominator of Eq. (\ref{tan2alpha}) is compensated by the 
radiative correction $\delta M^2_{11}$ or $\delta M^2_{22}$ or both,
$\alpha$ will largely diviates from $\beta - \frac{\pi}{2}$ (e.g., $\alpha\sim 0.5$) and consequently
$\sin(\beta-\alpha)$ can be lifted from $1$.

\section{E\lowercase{xperimental constraints of
 $(g-2)_{\mu}$ and $b\rightarrow s \gamma$}}
In this section we analyze the experimental constraints from $(g-2)_{\mu}$ and $b\rightarrow
 s \gamma$ which will be imposed on the parameter space of MSSM in our numerical analysis.

The muon anomalous magnetic moment $a_{\mu} \equiv \frac{1}{2}(g-2)_{\mu}$ constraint from
the recent Brookhaven E821 experiment~\cite{E821} gives a $2\sigma$ bound
on the supersymmetry contribution, 
\begin{equation}
11\times 10^{-10}\,<\,  a^{SUSY}_{\mu} \,< \, 75 \times 10^{-10}.
\end{equation}

The one loop supersymmetric contributions to $a_{\mu}$ come from the diagrams with 
chargino-sneutrino in the loop and neutralino-smuon in the loop respectively~\cite{martin}.
The chargino contribution is given in the limit
of $|M_2| \ll |\mu|$,
\begin{equation}
\delta a_{\mu}^{\tilde{\chi}^\pm} = 1.5 \tan\beta
(\frac{200{\rm GeV}}{M_{\tilde{\nu}}})^2 (\frac{M_2}{100{\rm GeV}})
(\frac{1000{\rm GeV}}{\mu}) (\frac{F_2^C(\frac{M_2^2}{M^2_{\tilde{\nu}}})}{3})
10^{-10},
\label{mug-2chargino}
\end{equation}
where $F^C_2(x)$ is defined in ~\cite{martin}, with
$F^C_2(1)=1$, $F^C_2(0.25)=2.5$.
When $\tan\beta$ is large,
it can contribute within the experimental bound.
The neutralino contribution can be important from Bino-
smuon loop, as emphasized by Martin and Wells in ref.\cite{martin}.
In the large $\mu$
(say $\mu \sim 1$Tev) limit,
$|M_1| \ll |M_2|,|\mu|$ and two smuon are nearly degenerate
$m_{\tilde{\mu}_1} \simeq m_{\tilde{\mu}_2}$. The Bino contribution can be
written as
\begin{equation}
\delta a_{\mu}^{\rm B-ino} = 3.0 \tan\beta
(\frac{M_1}{100{\rm GeV}})
(\frac{\mu - A_{\mu}\cot\beta}{1000{\rm GeV}})
(\frac{200{\rm GeV}}{M_{\tilde{\mu}}})^4
(\frac{F_2^{N \prime}(\frac{M_1^2}{M_{\tilde{\mu}}})}{3}) 10^{-10}
\label{mug-2neutra}
\end{equation}
where $F_2^{N \prime}(x)$ is defined as $F_2^{N \prime}(x) =
2F_2^N(x) + 2x dF_2^N(x)/dx$, with $F_2^{N \prime}(1) = 1$ and
$F_2^{N \prime}(0.25) = 2.4$. The
$(\mu - A_\mu\cot\beta)\tan\beta$ term in Eq. (\ref{mug-2neutra}) comes from smuon mixing.
It is obvious that when $\mu\tan\beta$ is large (say, $\mu\tan\beta \sim 20$Tev)
 and $M_{\tilde{\mu}}$ is not too
large, the experimental constraint of $(g-2)_\mu$ can be satisfied
by the Bino contribution.

The $b \rightarrow s \gamma$ decay branch ratio from CLEO~\cite{cleo99}
is given as
\begin{equation}
2\times 10^{-4} < BR_{exp}(b\rightarrow s\gamma) < 4.2 \times 10^{-4}.
\end{equation}
In our numerical analysis, we use the leading order calculation of
BR($b\rightarrow s\gamma$) because of the lack of full next-to-leading
order calculations in MSSM,
and additional $\pm 30\%$ uncertainty of the leading order calculation of
$b \rightarrow s \gamma$ has been considered. In this paper, since $M_{H^\pm}$
($M_{H^\pm} \sim 300$GeV) is not too large, it can enhance the
$BR(b\rightarrow s\gamma)$ significantly because of the constructive
contributions of the charged Higgs $H^\pm$ and
Standard Model (SM) charged gauge boson $W^\pm$.
This gives a serious constraint on the supersymmetric contributions such that
they must be destructively interferent with those in SM. 
In supergravity models with large $\tan\beta$, the
chargino contribution is correlated to the sign of the product of
$\mu$ and $A_t$. For positive $\mu A_t$, the chargino contribution interferes
with that in SM constructively, and for negative $\mu A_t$,
destructively~\cite{carena94}.

\section{T\lowercase{he} L\lowercase{ightest}
H\lowercase{iggs boson production at
$\gamma\gamma$ colliders}}
We follow the formula given in ~\cite{h2gamma}~\cite{higgshunter}.
The $\gamma\gamma$
width of the lightest Higgs is expressed as
\begin{equation}
\Gamma(h\rightarrow \gamma \gamma) = \frac{\alpha^2 g_2^2}{1024 \pi^3}
\frac{m_h^3}{M_w^2} |\sum_{i}I_h^i|^2
\end{equation}
where $i$ runs over all the loop contributions. The different kinds of $I_h^i$
are given as
\begin{eqnarray}
I_h^f &=& N_{cf}Q_f^2 R_f F_{1/2}(\tau_f),\\
I_h^w &=& R_w F_1(\tau_w),\\
I_h^{H^\pm} &=& R_{H^\pm} \frac{M_W^2}{M_{H^\pm}^2} F_0(\tau_{H^\pm}),\\
I_h^{\tilde{f}} &=& N_{cf} Q_f^2 R_{\tilde{f}} \frac{M_Z^2}{M_{\tilde{f}}}
F_0(\tau_{\tilde{f}}),\\
I_h^{\chi^\pm} &=& R_{\chi^\pm} \frac{M_W}{M_{\chi^\pm}}
F_{1/2}(\tau_{\chi^\pm})
\end{eqnarray}
where $\tau_i = 4 M_i^2/M_h^2$, $N_c=3$ for quarks and squarks, $N_c=1$ for
leptons and sleptons, the loop functions $F_{1/2}(x),F_{1}(x),F_{0}(x)$
are given in Ref.~\cite{h2gamma}~\cite{higgshunter},
and $R_i$ are defined as follow
\begin{eqnarray}
R_{u,c,t} &=& \frac{\cos\alpha}{\sin\beta},\\
R_{d,s,b,e,\mu,\tau} &=& - \frac{\sin\alpha}{\cos\beta},\\
R_W &=& \sin(\beta - \alpha),\\
R_{H^\pm} &=& \sin(\beta-\alpha)+
\frac{\cos 2\beta\sin(\beta+\alpha)}{2\cos^2\theta_W},\\
R_{\tilde{f}_{L,R}} &=& \frac{M_f^2}{M_Z^2} R_f \mp
(I_f^3 - Q_f\sin^2\theta_W)\sin(\beta+\alpha), \label{LR}\\
R_{\tilde{\chi_i^\pm}} &=& 2(S_{ii}\cos\alpha - Q_{ii}\sin\alpha)
\end{eqnarray}

If $\tan\beta$ is large, the bottom Yukawa coupling receives the large correction,
the coupling $R_b$ is
\begin{equation}
R_b \simeq - \frac{\sin\alpha}{\cos\beta}
\frac{1}{1+\Delta_b}( 1 - \frac{\Delta_b}{\tan\alpha \tan\beta}).
\label{rb}
\end{equation}
In Eq. (\ref{rb}) $\Delta_b$ is given at one-loop as~\cite{deltab}
\begin{equation}
\Delta_b \simeq \frac{2\alpha_s}{3\pi} M_{\tilde{g}}\mu \tan\beta
I(M_{\tilde{b}_1},M_{\tilde{b}_2},M_{\tilde{g}}) +
\frac{Y_t}{4\pi} A_t\mu \tan\beta I(M_{\tilde{t}_1},M_{\tilde{t}_2},\mu)
\end{equation}
where $\alpha_s=g^2_s/4\pi$ and $Y_t=h_t^2/4\pi$, $h_t$ is top Yukawa coupling and the function $I$ is defined by
\begin{equation}
I(x,y,z) =\frac{x^2 y^2\ln(x^2/y^2)+y^2 z^2\ln(y^2/z^2)+z^2 x^2\ln(z^2/x^2)}
{(x^2 - y^2) (y^2 - z^2) (x^2 - z^2)}
\end{equation}

Since the off-diagonal elements of the mass-squared matrix of the third generation sfermion 
are large, which is considered in this paper, left- and right-handed sfermion mixing should
be included in the Higgs sfermion couplings. In stead of 
Eq. (\ref{LR}), one has
\begin{eqnarray}
\label{rup1}
R_{\tilde{f}_1} &=& \frac{M_f^2}{M_Z^2}R_f -
(I_f^3\cos^2\theta_{\tilde{f}} - Q_f\sin^2\theta_w\cos 2\theta_{\tilde{f}})
\sin(\beta+\alpha) - \frac{M_f\sin 2\theta_{\tilde{f}}}{2M_Z^2\sin\beta}
(A_f\cos\alpha - \mu\sin\alpha),\\
\label{rup2}
R_{\tilde{f}_2} &=& \frac{M_f^2}{M_Z^2}R_f -
(I_f^3\sin^2\theta_{\tilde{f}} + Q_f\sin^2\theta_w\cos 2\theta_{\tilde{f}})
\sin(\beta+\alpha) + \frac{M_f\sin 2\theta_{\tilde{f}}}{2M_Z^2\sin\beta}
(A_f\cos\alpha - \mu\sin\alpha)
\end{eqnarray}
 for the top squark and 
\begin{eqnarray}
R_{\tilde{f}_1} &=& \frac{M_f^2}{M_Z^2}R_f -
(I_f^3\cos^2\theta_{\tilde{f}} - Q_f\sin^2\theta_w\cos 2\theta_{\tilde{f}})
\sin(\beta+\alpha) - \frac{M_f\sin 2\theta_{\tilde{f}}}{2M_Z^2\sin\beta}
(A_f\sin\alpha - \mu\cos\alpha),\\
R_{\tilde{f}_2} &=& \frac{M_f^2}{M_Z^2}R_f -
(I_f^3\sin^2\theta_{\tilde{f}} + Q_f\sin^2\theta_w\cos 2\theta_{\tilde{f}})
\sin(\beta+\alpha) + \frac{M_f\sin 2\theta_{\tilde{f}}}{2M_Z^2\sin\beta}
(A_f\sin\alpha - \mu\cos\alpha)
\end{eqnarray}
for the bottom squark and tau slepton, where $\theta_{\tilde{f}}$ is the sfemion mixing angle.

The production cross section of the lightest Higgs boson at $\gamma\gamma$ colliders is
given as
\begin{eqnarray}
\sigma(\gamma\gamma\rightarrow h) &=& \frac{8\pi^2}{M_h^3}
\Gamma(h\rightarrow \gamma\gamma)\delta(1-\frac{\sqrt{s}}{M_h^2})\nonumber\\
 &=& \sigma_0 \delta(1-\frac{\sqrt{s}}{M_h^2})
\end{eqnarray}
with
\begin{eqnarray}
\sigma_0 &=& \frac{8\pi^2}{M_h^3}
\Gamma(h\rightarrow \gamma\gamma)
\end{eqnarray} 
where $\sqrt{s}$ is the energy of mass center.

\section{N\lowercase{umerical analyses}}
In our numerical work, for simplicity,
we assume the universal soft sypersymmetry
breaking squark mass $M_{\tilde{q}} = M_{\tilde{t}_R}
= M_{\tilde{b}_R} = M_{\tilde{c}_R} = M_{\tilde{s}_R} = M_{\tilde{u}_R} =
M_{\tilde{d}_R}$, and universal soft sypersymmetry breaking
slepton mass $M_{\tilde{l}} = M_{\tilde{\tau}_R}= M_{\tilde{\mu}_R}$
which is different from the squark sector.
With the assumption of the universal slepton mass,
the requirement that the stau mass should be positive puts a severe constraint on
smuon mass matrix elements, because the only difference between the stau and smuon
mass matrices is just that $M_{\tau}$ in the stau mass matrix
is replaced by $M_{\mu}$ in the smuon mass matrix.
With $M^2_{\tilde{l}} > M_\tau\tan\beta |\mu - A_\tau\cot\beta|$
and $M_\tau/M_\mu \simeq 17$, the two smuon mass are nearly degenerate,
but smuon mixing angle is nonzero in the limit of large $\mu\tan\beta$.
We analyze the parameter region under experimental constraints from lower
bounds of sypersymmetry particles and Higgs bosons~\cite{pdg2000},
$b\rightarrow s\gamma$ and the muon anomalous magnetic moment $(g-2)$
in three cases:
$(a)$low $\tan\beta$, $(b)$moderate $\tan\beta$,
$(c)$large $\tan\beta$.
In all our analyses and numerical work, we take $A_\mu = A_\tau = 0$,
because of large $\mu\tan\beta$ in the off-diagonal mass matrix elements
$(A_{\mu,\tau} - \mu\tan\beta)$.
We also fix $M_{\tilde{g}} = 1$TeV. This parameter only appears in the
bottom Yukawa correction $\Delta_b$.
The radiative corrections to the CP-even Higgs mass matrix element
$\delta M^2_{ij} (i,j=1,2)$ depend on top and bottom squark mass matrix
parameters $M_{\tilde{q}}$, $\mu$, $A_t$ and $A_b$.
The neutralino as the lightest supersummetry particle(LSP) is also assumed.

\begin{figure}
\begin{tabular}{cc}
\epsfxsize 8cm
\epsfysize 8cm
\epsffile{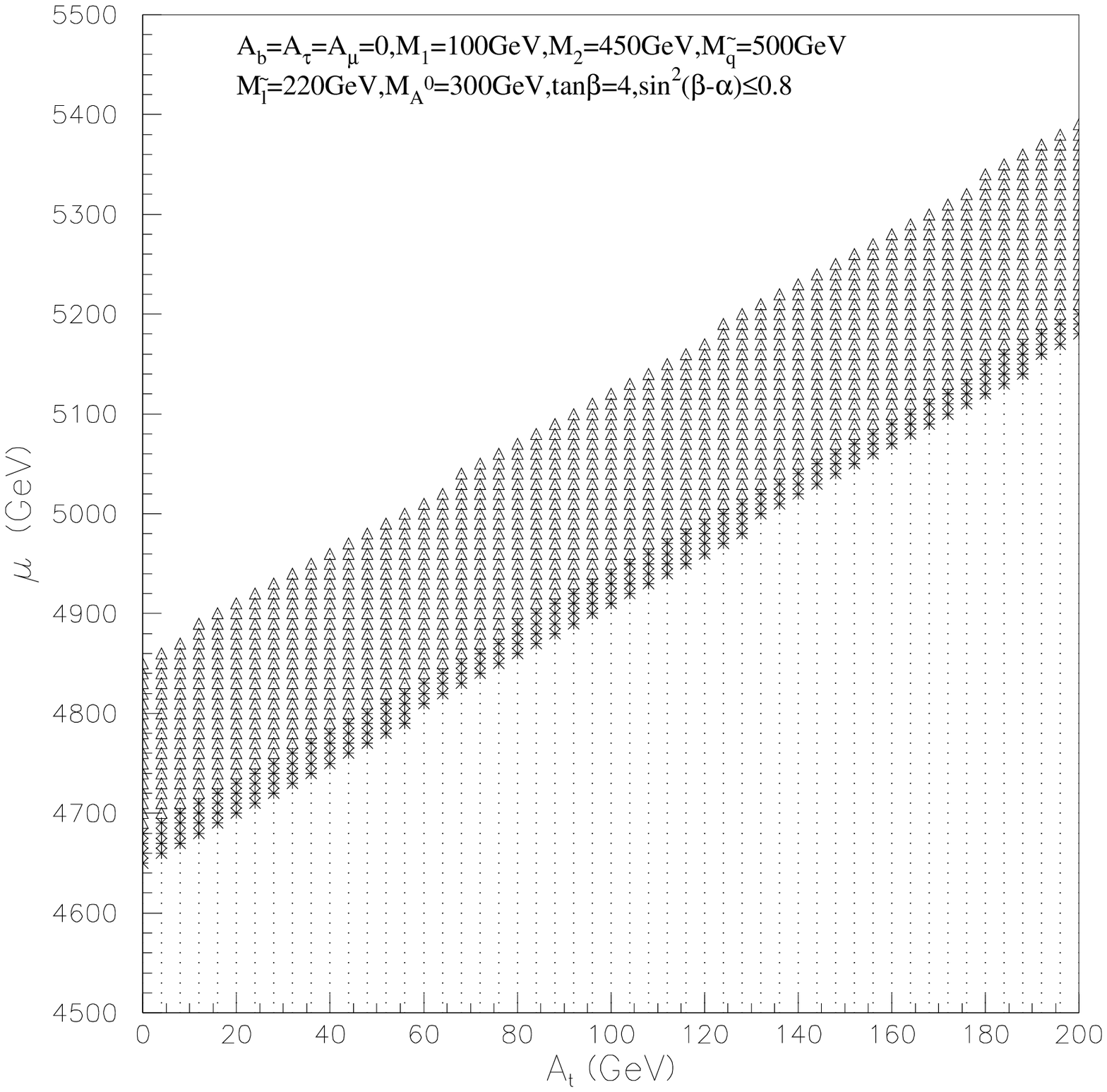}&
\epsfxsize 8cm
\epsfysize 8cm
\epsffile{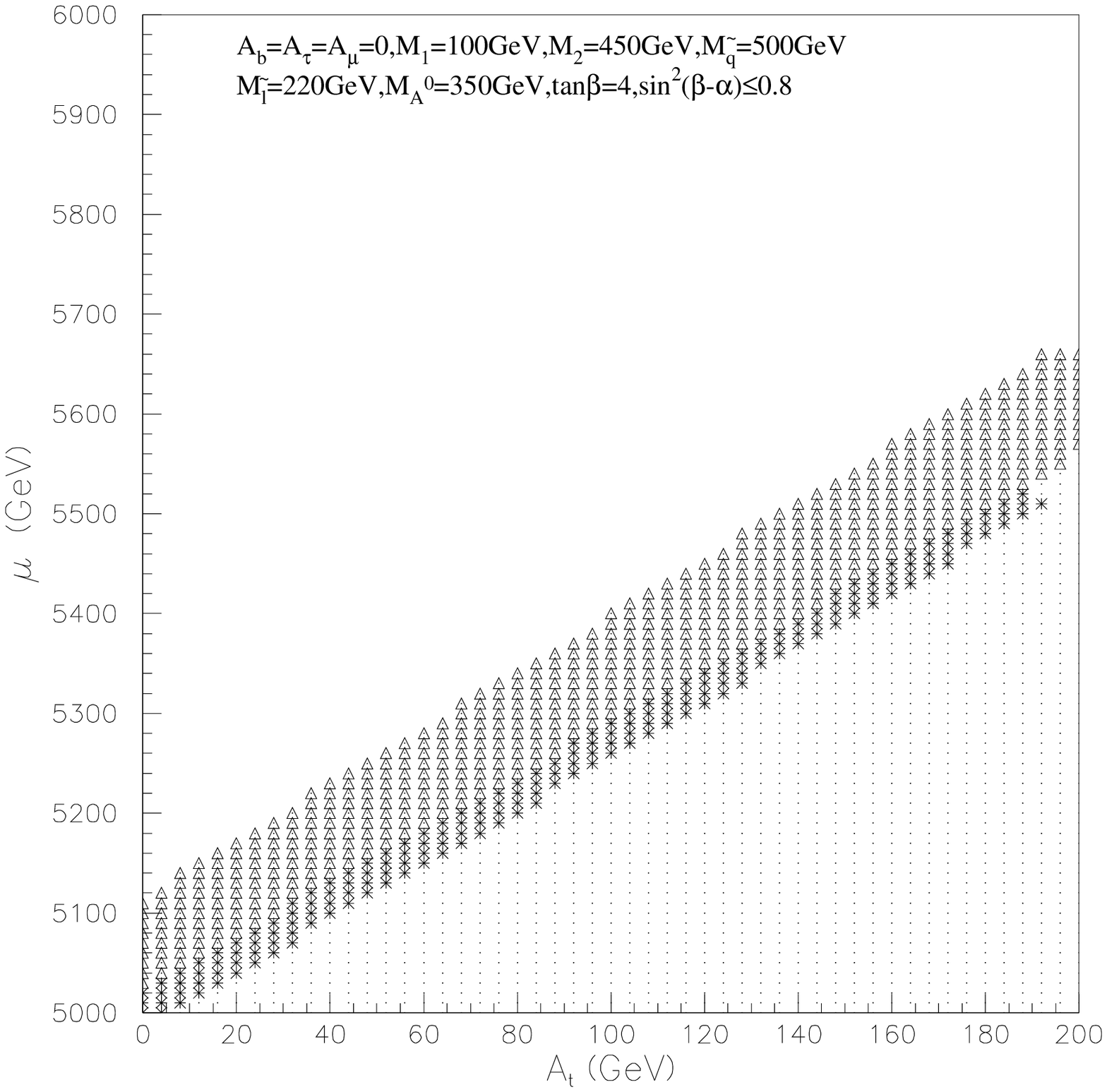}\\
A&B\\
\epsfxsize 8cm
\epsfysize 8cm
\epsffile{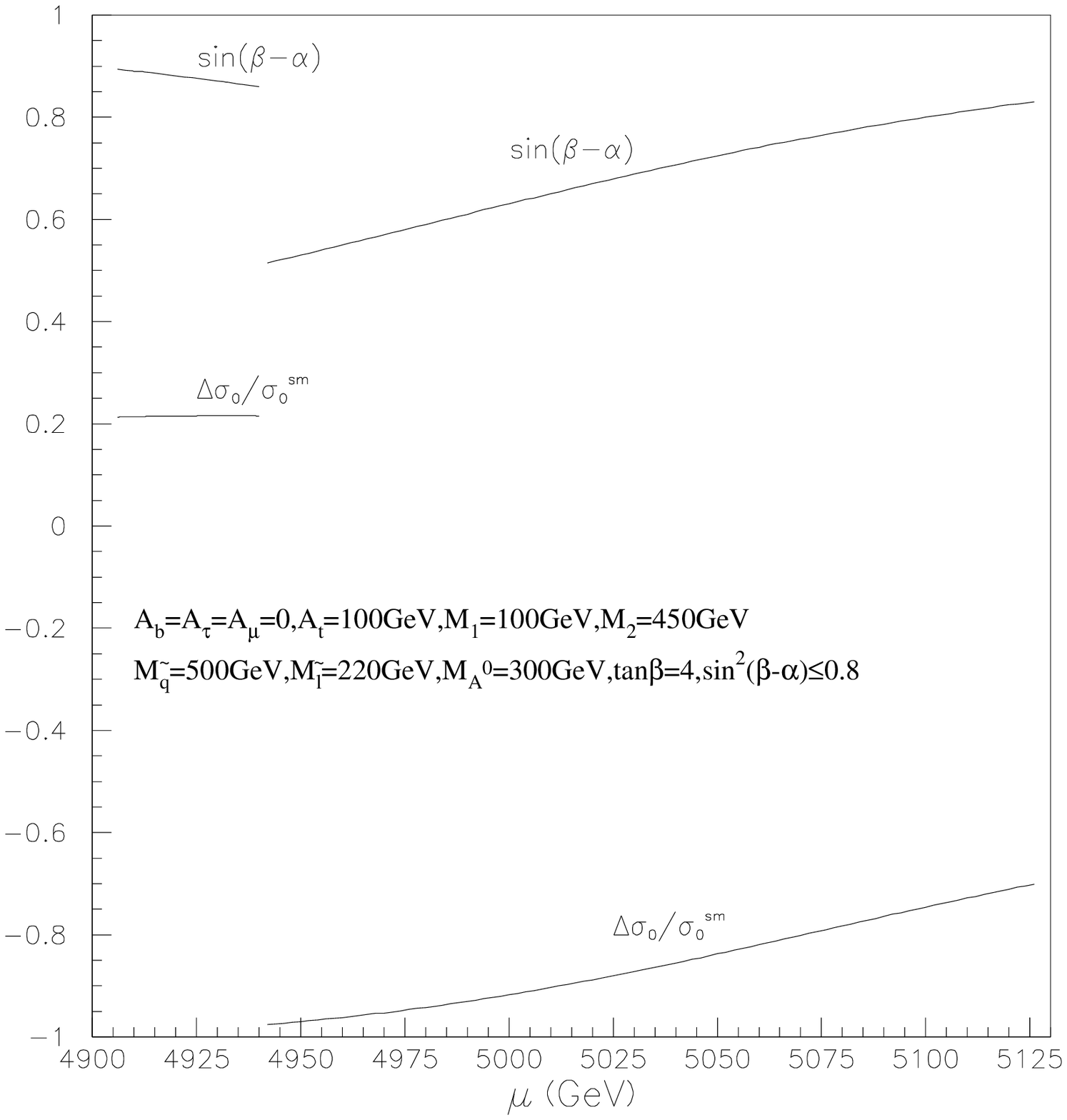}&
\epsfxsize 8cm
\epsfysize 8cm
\epsffile{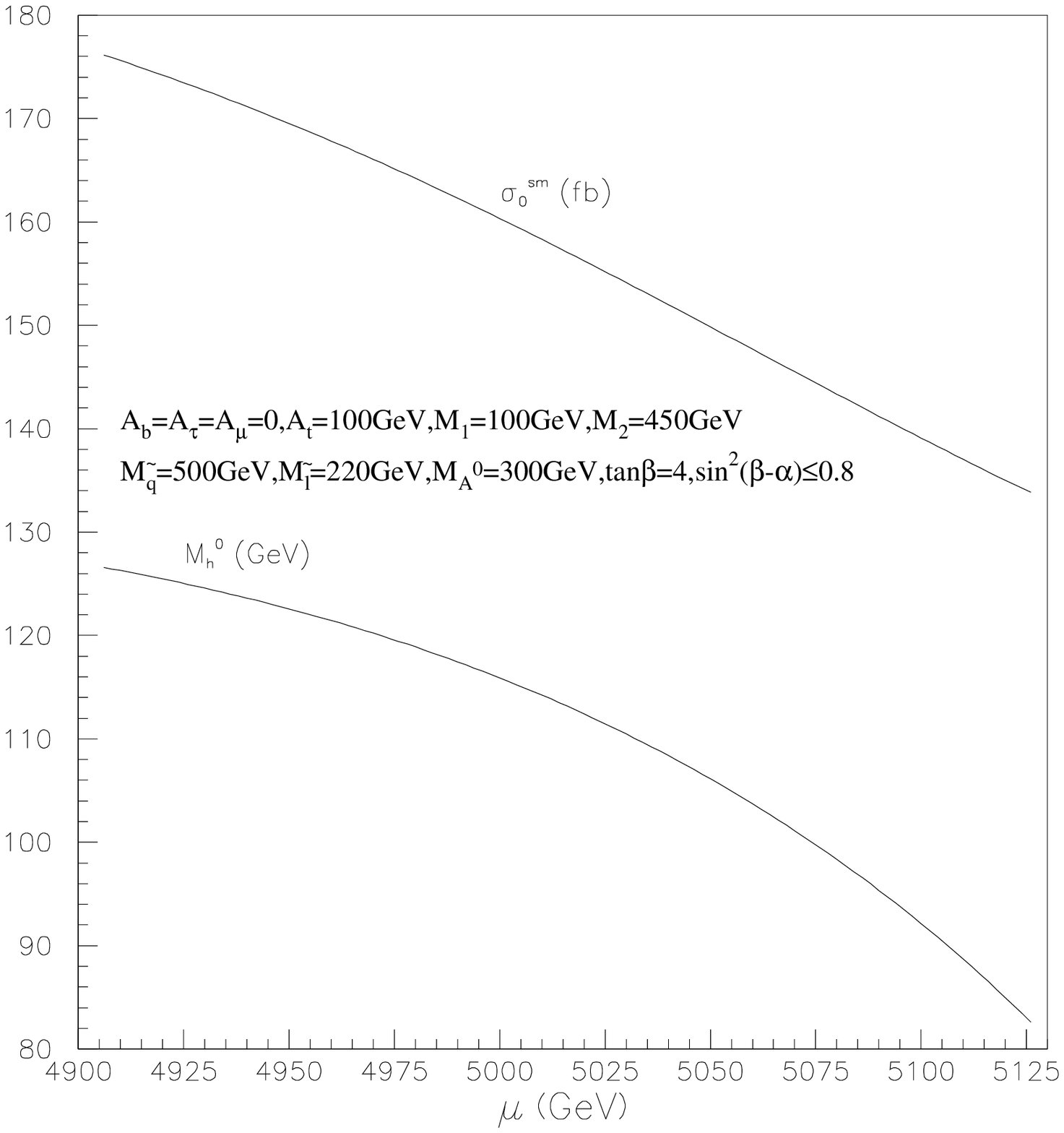}\\
C&D
\end{tabular}
\caption{The low $\tan\beta=4$ case. The dotted, triagle and star regions are
allowed by experiments.
The star (triangle) region corresponds to $\sin^2(\beta-\alpha) \le 0.8$ for
negative (positive) $\alpha$.
}
\label{figurelow}
\end{figure}

First, we work in the small $\tan\beta$ case, say $\tan\beta=4$.
Compared with the top Yukawa coupling, the bottom Yukawa coupling is small,
and $A_b=0$ is taken.
We fix $M_{\tilde{q}}$, $M_{\tilde{l}}$, $M_{A^0}$, $A_b$, $M_1$ and $M_2$,
all the triangle, star and dotted areas shown in Fig.~\ref{figurelow} are
experimentally allowed.
In this parameter region, $M_1 = 100$GeV and $M_2 = 450$GeV, the dominant
contribution to the muon anomalous magnetic moment $(g-2)$ comes from the neutralino-smuon loop.
In the region, $\mu \sim 5$TeV,  the chargino-sneutrino loop contribution is 
negligible as it can be seen from
Eq. (\ref{mug-2chargino}). Because of large chargino and stop masses and low
$\tan\beta$, the $b\rightarrow s\gamma$ can be satisfied easily.
When we further require $\sin^2(\beta-\alpha) \le 0.8$, the two regions denoted by
triangle and star appear, corresponding to the CP-even Higgs mixing angle
$\alpha > 0$ and $\alpha < 0$ respectively.
We see that in the two regions $\mu$ is in
the range between $9.3M_{\tilde{q}}$ and $10M_{\tilde{q}}$ and $A_t$
is within a small range near $0$.
When $M_{A^0}$ is raised, the allowed parameter region is decreased, as can be seen by
comparing the Fig. 1A and 1B.
As an illustration, we
present two points in Fig.~\ref{figurelow}$A$ as the cases $A$ and $B$ in
Tab.~\ref{table}, corresponding to the
positive and negative CP-even Higgs mixing angle $\alpha$ respectively.
We can see $\sqrt{|\delta M^2_{22}|}$ and $\sqrt{|\delta M^2_{12}|}$ are
the same order as $M_Z$, but $\delta M^2_{11}$ is about a factor of $6$ increase
of $\delta M^2_{22}$ in magnitude. With $\cos 2\beta \simeq -0.6$, $\sin 2\beta 
\simeq 2\cot\beta \simeq 0.4$, we notice that in the denominator of Eq. (\ref{tan2alpha}), the
$M^2_{A^0}$ term and $\delta M^2_{11}$ term compensate each other with an
order of magnitude larger than the numerator. This fine-tune gives the value
of $|\alpha| \sim 0.75$, and reduces $\sin^2(\beta-\alpha)$. With this picture in
mind, we can understand that when $M_{A^0}$ increases, the fine-tune of
$M^2_{A^0}$ and $\delta M^2_{11}$ terms in the denominator of Eq. (\ref{tan2alpha})
becomes more difficult, and as shown Fig.~\ref{figurelow}$B$, we get a parameter region 
significantly smaller than that in Fig.~\ref{figurelow}$A$.

\begin{figure}
\begin{tabular}{cc}
\epsfxsize 8cm
\epsfysize 8cm
\epsffile{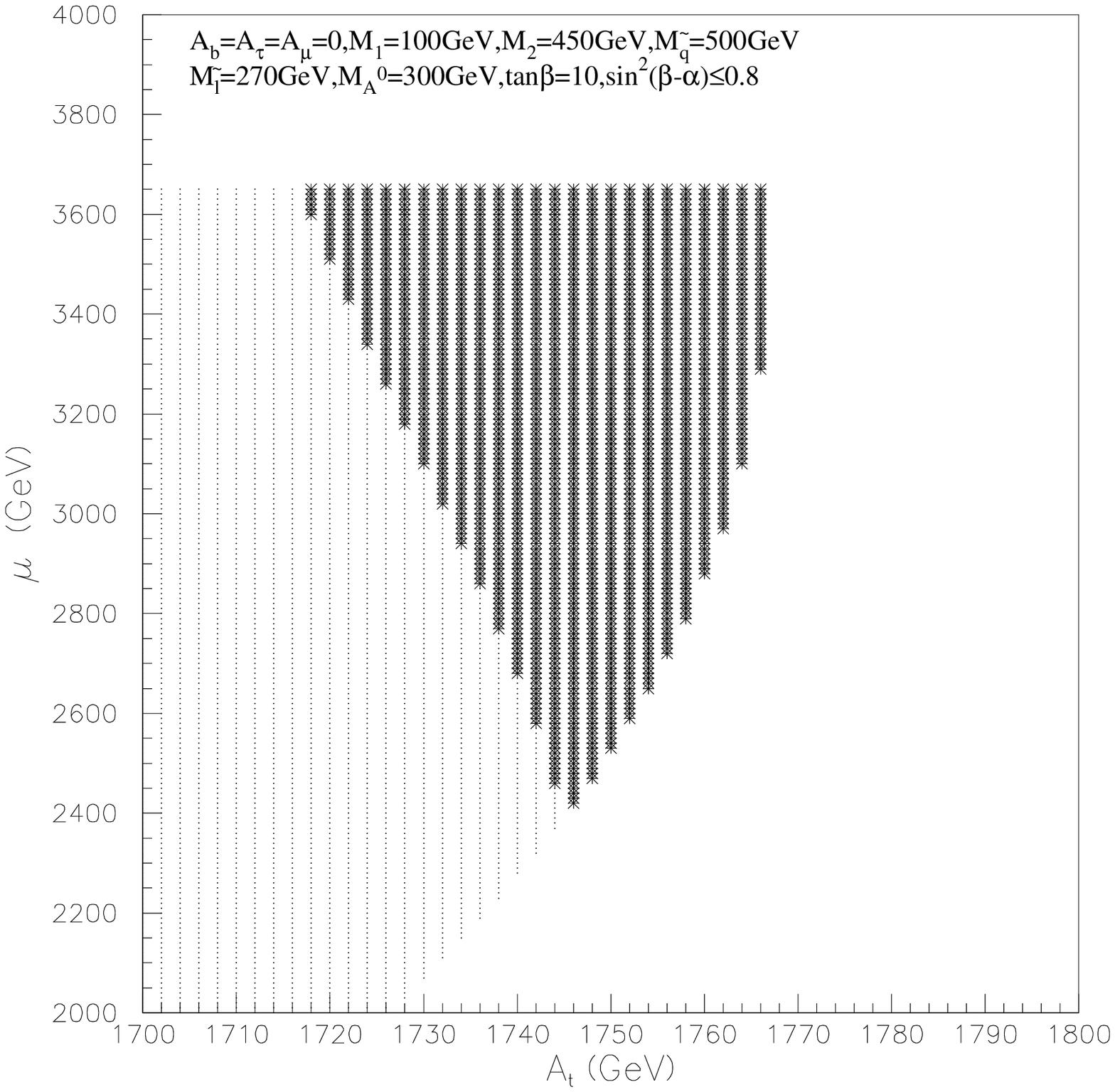}&
\epsfxsize 8cm
\epsfysize 8cm
\epsffile{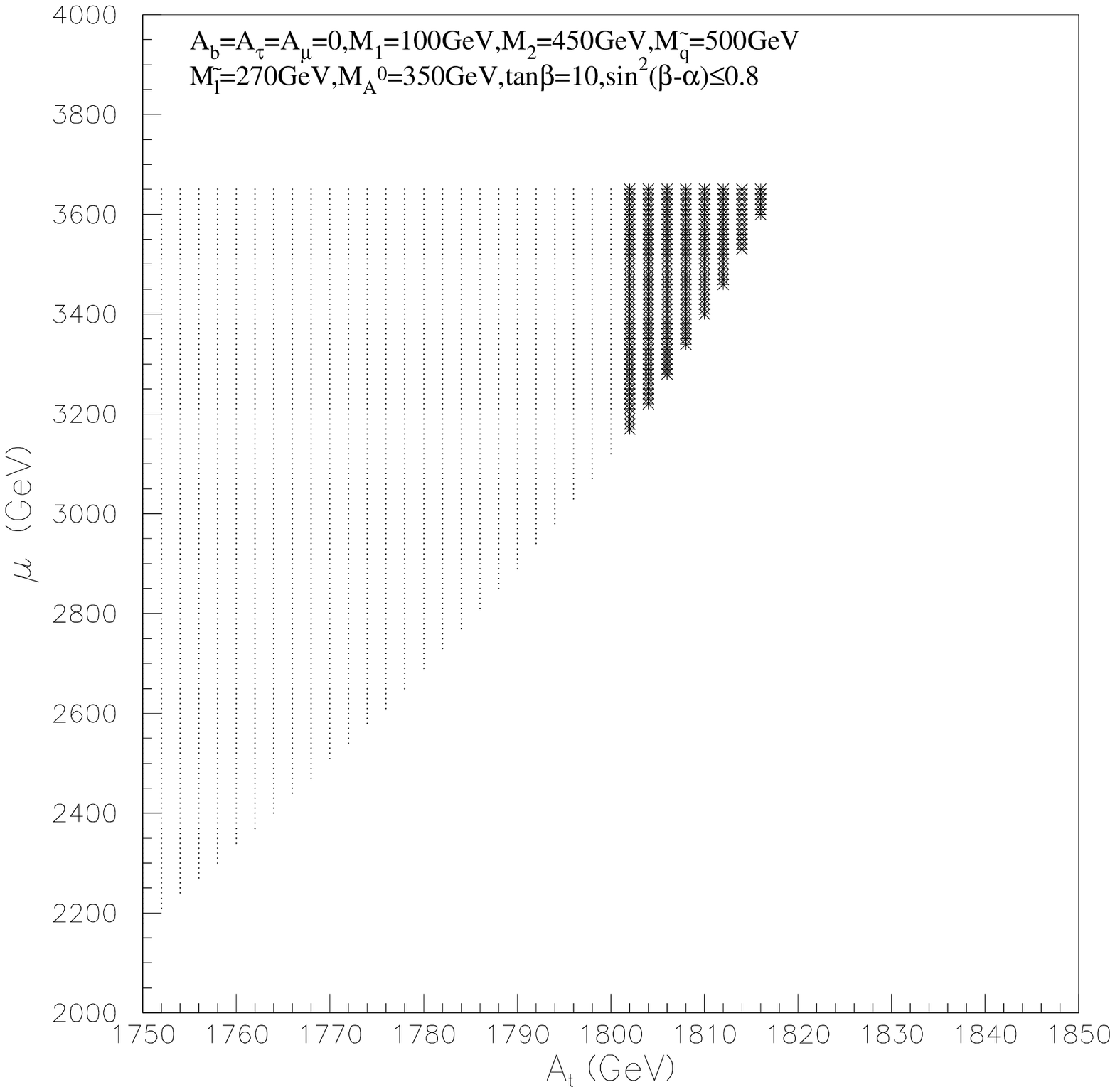}\\
A&B\\
\epsfxsize 8cm
\epsfysize 8cm
\epsffile{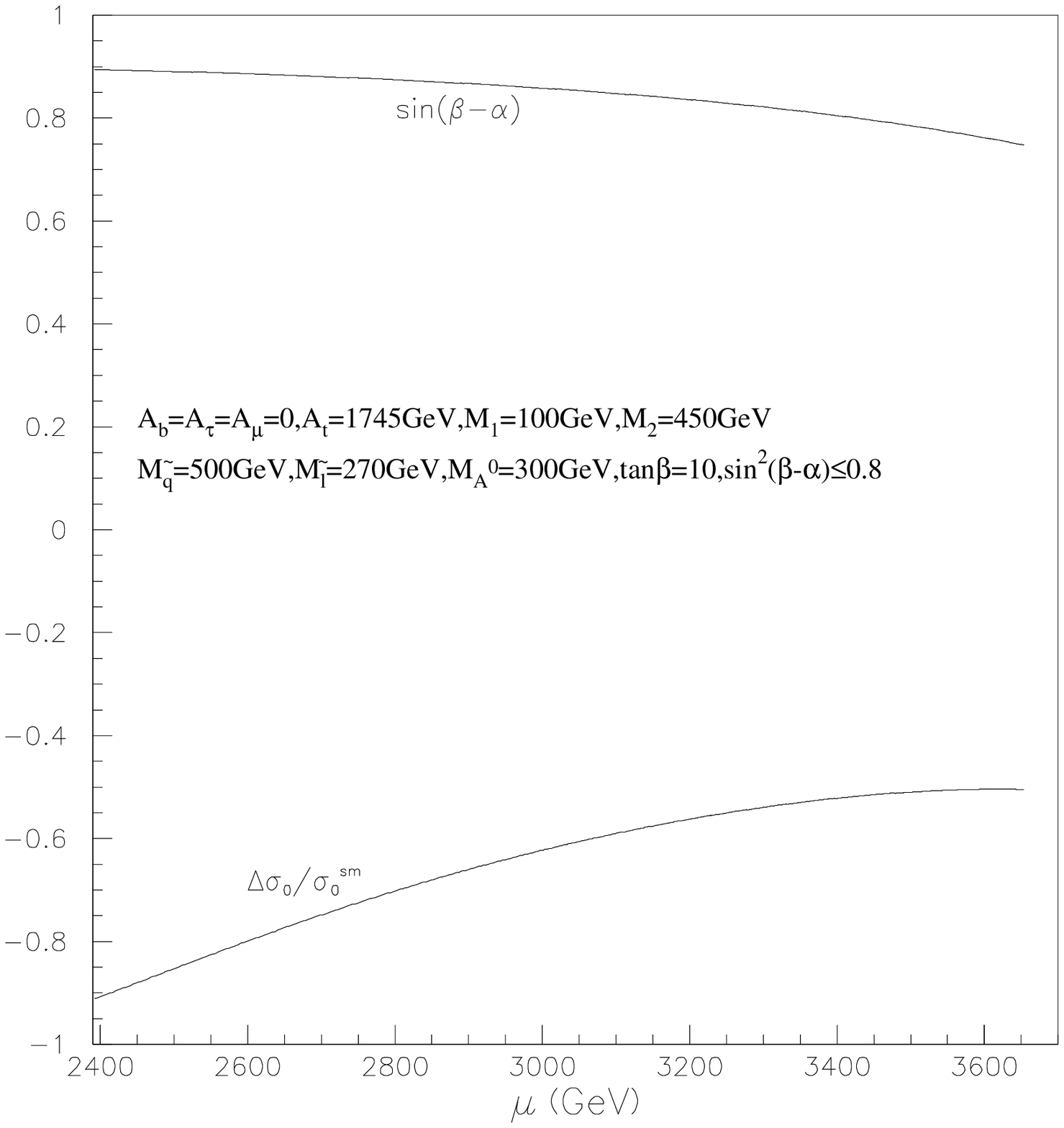}&
\epsfxsize 8cm
\epsfysize 8cm
\epsffile{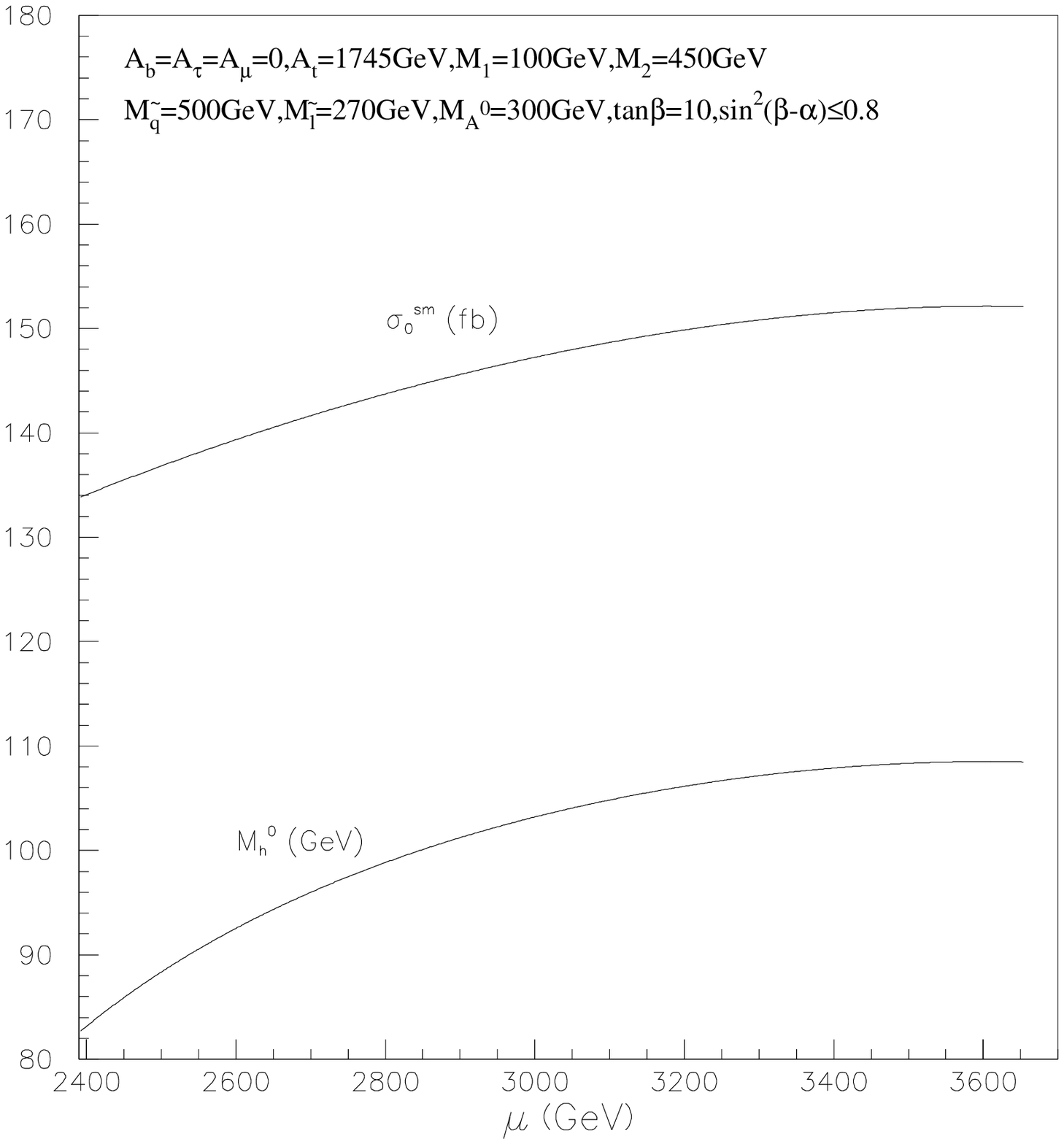}\\
C&D
\end{tabular}
\caption{The moderate $\tan\beta=10$ case. The dotted and star regions are allowed by
experiments, and the star region corresponds to $\sin^2(\beta-\alpha) \le 0.8$.
}
\label{figuremod}
\end{figure}

Second, in the case of moderate $\tan\beta = 10$, we fix $M_{\tilde{q}}$,
$M_{\tilde{l}}$, $M_{A^0}$, $A_b$, $M_1$ and $M_2$, and the experimentally
allowed area is denoted by the dotted and star areas as shown in
Fig.~\ref{figuremod}. In this parameter region, because $\mu \simeq 3$TeV and
$\tan\beta=10$, the chargino contribution to the muon anomalous magnetic moment $(g-2)$ is not
large enough to rest in the experimental bound.
The dominant contribution to $a_{\mu}$ comes from the neutralino-
sneutrino loop as seen from Eq. (\ref{mug-2neutra}).
In Fig.~\ref{figuremod} the star area corresponds to the region of the parameter space where
 $\sin^2(\beta-\alpha) \le 0.8$.
We see that $\mu$ is in the range between $4.8 M_{\tilde{q}}$ and
$7.3 M_{\tilde{q}}$ and $A_t$ is in a small range near $3.5 M_{\tilde{q}}$.
The fine-tune property of Eq. (\ref{tan2alpha}) is similar to that in the low $\tan\beta$
case. When $M_{A^0}$ increases, the allowed parameter region by experiments
and the $\sin^2(\beta-\alpha) \le 0.8$ requirement is minimized, as shown in
Fig.~\ref{figuremod}$B$.

\begin{figure}
\begin{tabular}{cc}
\epsfxsize 8cm
\epsfysize 8cm
\epsffile{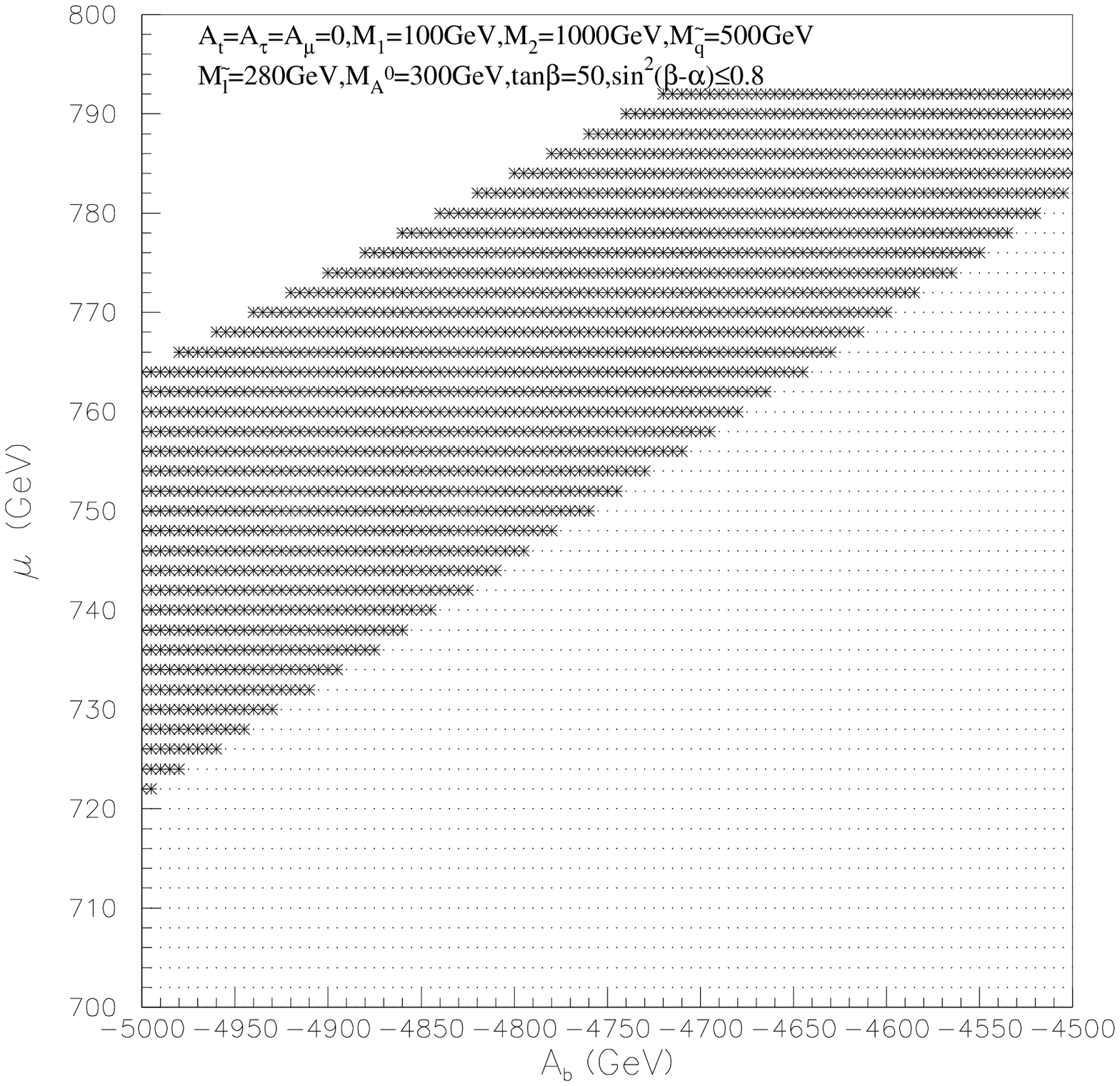}&
\epsfxsize 8cm
\epsfysize 8cm
\epsffile{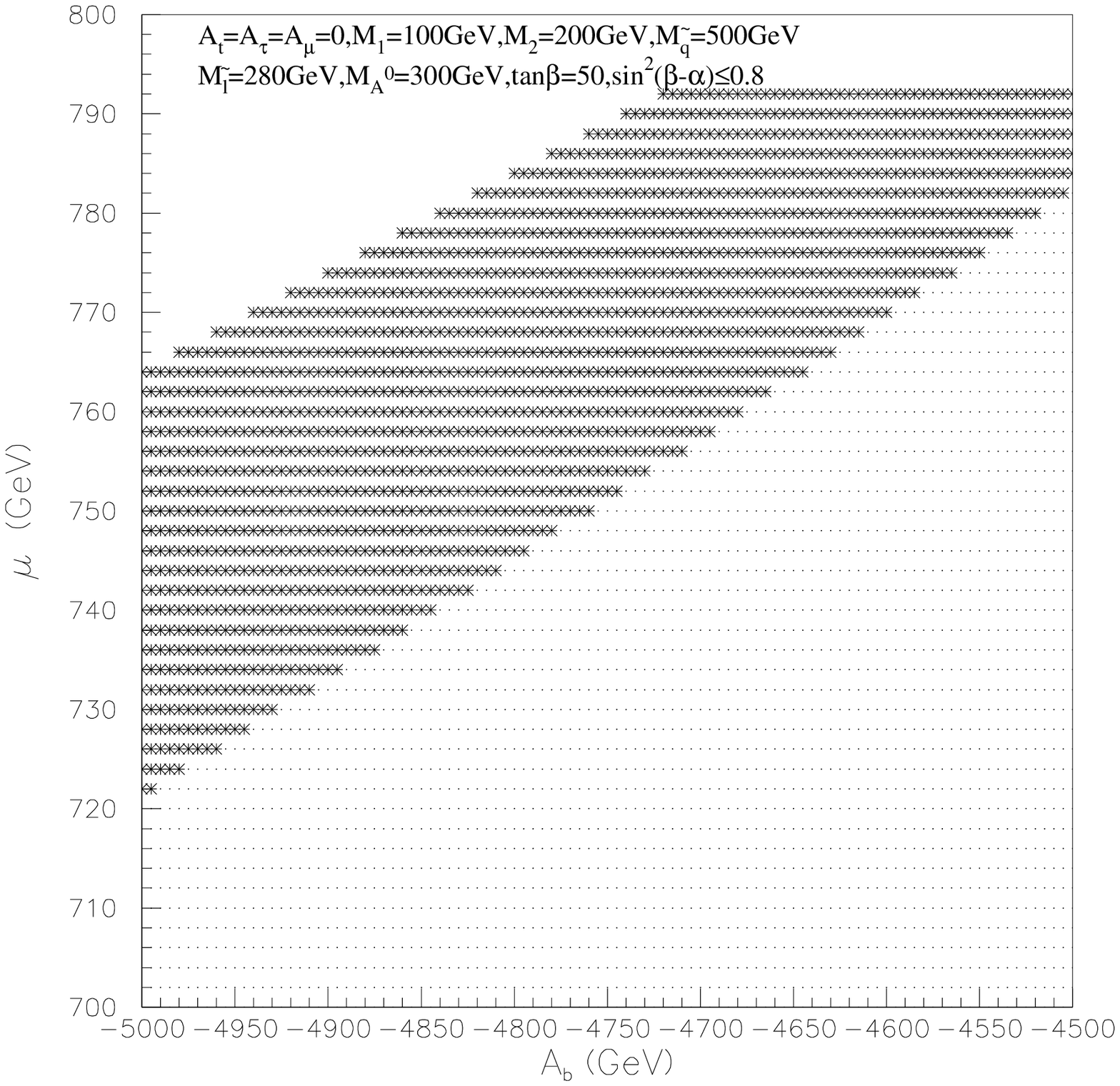}\\
A&B\\
\epsfxsize 8cm
\epsfysize 8cm
\epsffile{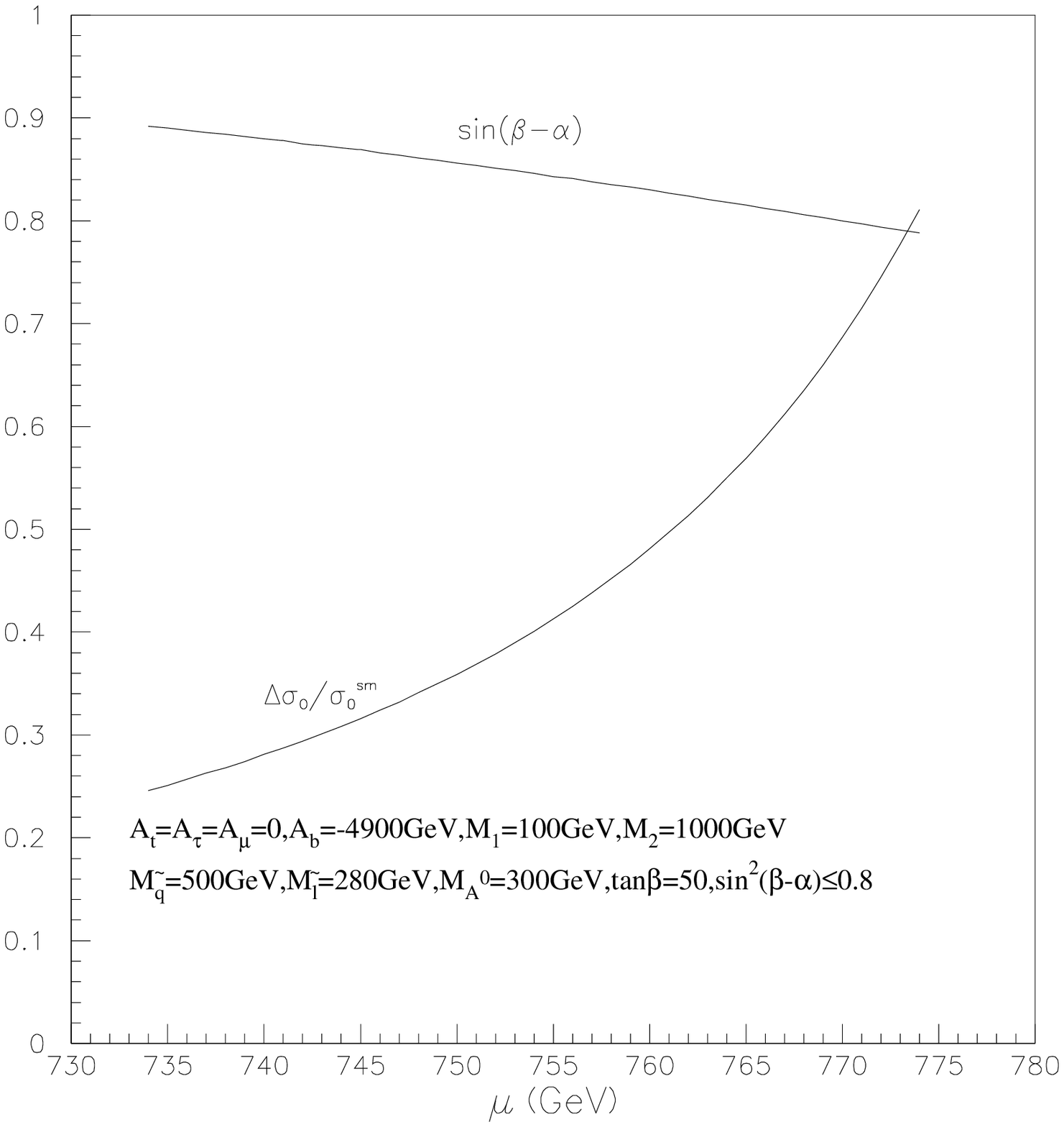}&
\epsfxsize 8cm
\epsfysize 8cm
\epsffile{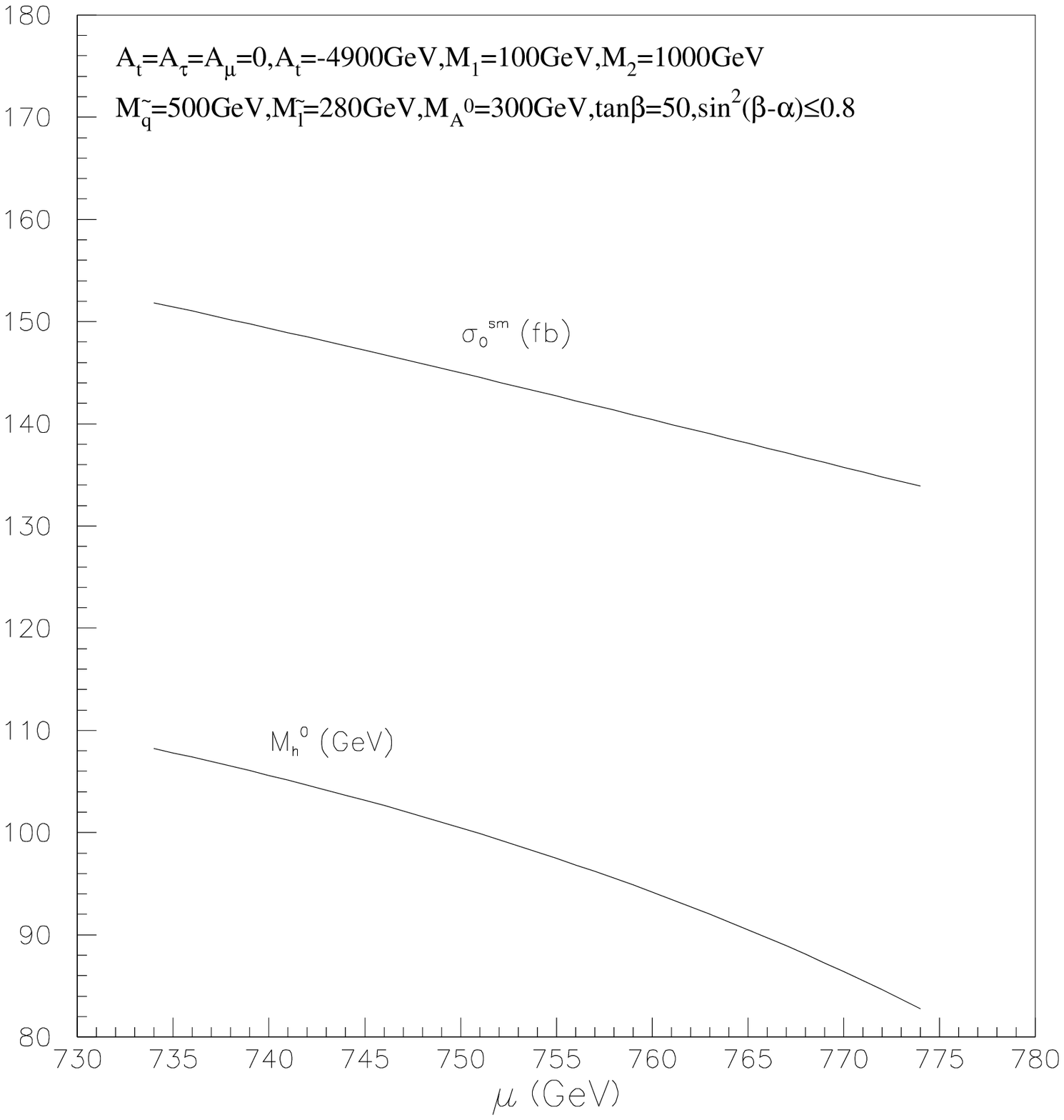}\\
C&D
\end{tabular}
\caption{The large $\tan\beta=50$ case. The dotted and star regions are allowed by
experiments, and the star region corresponds to $\sin^2(\beta-\alpha) \le 0.8$.
}
\label{figurelarge}
\end{figure}

Third, in the case of large $\tan\beta = 50$, since the bottom Yukawa coupling
is large compared with the top Yukawa coupling, we concentrate on the contributions
of the bottom and sbottom and take $A_t = 0$ for a while.
The experimentally allowed parameter region
of $\mu$ and $A_b$ is shown in Fig.~\ref{figurelarge}
in both the dotted and star areas, with fixed
$M_{\tilde{q}}$, $M_{\tilde{l}}$, $M_{A^0}$, $A_t$, $M_1$ and $M_2$.
We notice that $\mu$ is in a small range near $1.5 M_{\tilde{q}}$ because
of the large $M_b\tan\beta$ in the off-diagonal sbottom mixing term
$M_b(A_b - \mu\tan\beta)$ and $A_b$ is in the range between
$-9 M_{\tilde{q}}$ and $-10 M_{\tilde{q}}$.
The star area is allowed by the requirement of
$\sin^2(\beta-\alpha) \le 0.8$. With $\tan\beta=50$ and $\mu \simeq 750$GeV,
we can see from Eq. (\ref{mug-2neutra}), the neutralino-smuon loop alone can
generate the muon anomalous magnetic moment $(g-2)$ within the experiment bound as
shown in Fig.~\ref{figurelarge}$A$ with large $M_2$,
where the chargino contribution can be neglected. But because of
large $\tan\beta=50$ and not too large $\mu \simeq 750$GeV, the chargino-sneutrino loop
alone can also generate a value large enough to satisfy the experimental bound,
as can be seen from Eq. (\ref{mug-2chargino}). This case is shown in
Fig.~\ref{figurelarge}$B$, where we choose $M_1=100$GeV 
simply because of the requirement that the neutralino is the LSP.
Even in this lower chargino mass case, we need not worry about the
$b\rightarrow s\gamma$ bound, because of nearly degenerate scalar
up-type quark masses arising from the assumption of the universal soft supersymmetry breaking
squark mass, not too large $\mu$ and $A_t=0$.
When $A_t\neq 0$, for example, $A_t=150$GeV, and the other parameters are the same as those
shown in Fig.~\ref{figurelarge}$B$,
Br($b\rightarrow s\gamma$) reaches its upper bound, because of
positive $\mu A_t$ which leads to that the chargino contribution interferes
with that in SM constructively. Furthermore, we can show in our numerical analysis
that if we assign a negative value
to $A_t$ (say, $-500$GeV), the SUSY contributions interfere destructively with that in SM
so that the $b\rightarrow s\gamma$ constraint can be easily satisfied.

The damped $\sin^2(\beta-\alpha)$ has significant effects on the vertices
of $g_{h^0 W^+W^-}$, $g_{h^0ZZ}$ and $g_{h^0\gamma\gamma}$ and consequently on 
Higgs productions at photon colliders, as pointed out in Introduction.
With parameters chosen in the regions denoted by the star area in the figures, i. e.,
the regions allowed by experiments and the requirement of  $\sin^2(\beta-\alpha) \le 0.8$,
  we show our numerical result of the lightest CP-even Higgs production at
$\gamma\gamma$ colliders in Tab.~\ref{table}. Since the dominant contribution
to this process comes from $W^+W^-$ loop, with damped $\sin(\beta-\alpha)$,
this $W^+W^-$ loop contribution is reduced. For the case C in the table, because 
$|\alpha| \sim 0.5$, the Yukawa coupling of the lightest neutral
CP-even Higgs boson to top quarks is lifted from the decoupling case so that the top-antitop loop
 contribution is also reduced. Therefore, the production cross section is significantly reduced 
compared with that in SM. In large $\tan\beta$ limit, the bottom
Yukawa effect is enhanced, which corresponds to the case D in which the production section is 
significantly enhanced compared with that in SM due to $\alpha \sim -0.5$.  Since
the off-diagonal term of the stop mixing matrix is large in the cases $A$, $B$ and $C$,
as seen in Eqs. (\ref{rup1}), (\ref{rup2}),
$R_{\tilde{t}}$ enhances the stop contribution significantly.
The ratio $\Delta\sigma_0/\sigma_0^{sm} = (\sigma_0^{MSSM} -
\sigma_0^{sm})/\sigma_0^{sm}$ of the lightest Higgs boson production
cross section at photon colliders as a function of $\mu$ for different
$\tan\beta$ cases are shown  in
Fig.~\ref{figurelow}$C$,~\ref{figuremod}$C$,~\ref{figurelarge}$C$.
And additional quantities $\sin(\beta-\alpha)$,
$\sigma_0^{sm}$ and $M_{h^0}$ are also drawn in Figs. 1,2,3. Fig.~\ref{figurelow}$C$ is of a clear
manifest of the fine-tune property in Eq. (\ref{tan2alpha}) with the negative CP-even mixing
angle. We can see the ratio $\Delta\sigma_0/\sigma_0^{sm}$ is significantly different from zero
in all three cases of $\tan\beta$, with an either decreased or increased change
of the lightest Higgs boson production cross section compared with that in SM.
 
\section{C\lowercase{onclusions}}
In summary, through including the radiative corrections to the CP-even neutral Higgs sector from large
stop and sbottom off-diagonal matrix elements,
we have found allowed parameter regions at small, moderate and large $\tan\beta$
with damped $\sin^2(\beta-\alpha)$ of the Higgs couplings and $M^2_{A^0} \gg M^2_Z$ under the
experimental constraints of lower mass bounds of superparticles and
Higgs bosons, $b\rightarrow s\gamma$ and the recent  $(g-2)_{\mu}$. That is, the decoupling 
limit is not a necessary consequence of $M^2_{A^0} \gg M^2_Z$ and dependent of the other parameters
in MSSM, in particular, $\mu, \tan\beta$ and $A_{t,b}$, at least in the
 ranges of values of parameters which we consider in the paper,  after one includes the radiative
corrections to the Higgs sector. However, 
with increased $M_{A^0}$, the parameter regions are decreased. We find in all
three case that there are large deviations of the production cross section
of the lightest Higgs boson at photon colliders in MSSM to that in the Standard Model, 
with either decreased or increased results, which are expected to be tested at the future
photon colliders.
Finally, one can see from Tab.~\ref{table} that the mass splitting between $M_H^0$ and $M_A^0$ is
large, which is another implication of non-decoupling limit, as discussed in~\cite{akeroyd01} 
for the case of the large mass difference between the charged Higgs boson $H^\pm$ and heavy 
CP-even Higgs boson $H^0$, CP-odd Higgs boson $A^0$.
\section*{Acknowledgments}
This research was supported in part by the National Nature Science
Foundation of China.

\newpage
\begin{table}
\caption{Msss spectra unit (GeV), $\sigma_0$ in unit fb,
$M_{A^0} = 300$GeV, $M_1 = 100$GeV.}
\label{table}
\begin{tabular}{rrrrrrrr}
Case & $\tan\beta$/$M_2$/$\mu$ & $A_t$/$A_b$ & $M_{\tilde{q}}$/$M_{\tilde{l}}$ &
$M_{\tilde{t}_{1,2}}$ & $M_{\tilde{b}_{1,2}}$ & $M_{\tilde{\tau}_{1,2}}$ &
$M_{\tilde{\chi}^\pm_{1,2}}$\\
$M_{\tilde{\chi}^0_{1,2,3,4}}$ & $M_{h^0,H^0}$ &
($\delta M^2_{11}$/$\delta M^2_{22}$)&
($\delta M^2_{12}$/$\delta M^2_{22}$)& $\sqrt{\delta M^2_{22}}$ &
$\alpha$ & $\sin(\beta-\alpha)$ & $I_W$/$I_W^{sm}$ \\
$I_t$/$I_t^{sm}$ &
$I_b$/$I_b^{sm}$ & $I_{\tilde{t}}$ &
$I_{\tilde{b}}$ & $I_{\tilde{\tau}}$ & $\sigma_o$ &
($\Delta \sigma_0$/$\sigma_0^{sm}$)  & \\
\hline
A & 4/450/4900 & 72/0 & 500/220 & 277/693 & 411/578 & 124/291 & 449/4901 \\
99/449/4900/4901 & 119/171 & $-6.6$ & 1.6 & 98 & 0.72 & 0.57 & 4.7/8.1 \\
$-1.4$/$-1.8$& $-0.1$/0.0 & $-1.6$ & $-0.0$ & $-0.1$ & 7.7 & $-0.95$ &\\
\hline
B & 4/450/4800 & 52/0 & 500/220 & 279/692 & 413/576 & 126/290 & 449/4801\\
99/449/4800/4801 & 123/178 & $-6.2$ & 1.5 & 99 & $-0.75$ & 0.88 & 7.3/8.3 \\
$-1.4$/$-1.8$ & 0.1/0.0 & 1.3 & $-0.0$ & $-0.1$ & 209.5 & 0.22 & \\
\hline
C & 10/450/3200 & 1750/0 & 500/270 & 168/727 & 342/621 & 134/363 & 449/3202\\
99/449/3200/3201 & 101/178 & $-9.8$ & 3.0 & 80 & 0.52 & 0.82 & 6.3/7.7 \\
$-1.6$/$-1.8$ & $-0.2$/0.0 & $-0.4$ & $-0.0$ & $-0.1$ & 64.0 & $-0.56$ & \\
\hline
D & 50/1000/760 & 0/$-4800$ & 500/280 & 525/530 & 268/657 & 113/384 & 748/1015\\
99/748/762/1015 & 100/187 & $-7.2$ & $-1.1$ & 92 & $-0.55$ & 0.86 & 6.7/7.7 \\
$-1.5$/$-1.8$ & 0.4/0.0 & $-0.1$ & 0.0 & $-0.1$ & 198.7 & 0.37 & 
\end{tabular}
\end{table}


\newpage

\newpage

\newpage


\end{document}